\documentclass[journal,final,a4paper]{IEEEtran}
\usepackage{bm}
\usepackage{bm,color,soul}
\usepackage{amsmath, amsthm, amssymb}
\usepackage[makeroom]{cancel}
\usepackage{xcolor}

\usepackage[pdftex]{graphics}
\usepackage{balance}
\theoremstyle{definition}

\usepackage{xspace}
\usepackage{amssymb}
\usepackage{graphicx}
\usepackage{cite}

\usepackage{bbm}
\usepackage{xcolor}
\usepackage{setspace}

\usepackage[normalem]{ulem}
\usepackage[user]{zref}
\newcommand{\sot}[1]{} 

\newcounter{revc}
\makeatletter \zref@newprop{revcontent}{} \zref@addprop{main}{revcontent}
\zref@newprop{revsec}{} \zref@addprop{main}{revsec}
\zref@newprop{revpage}{} \zref@addprop{main}{revpage}

\newcommand{\revi}[2]{
\zref@setcurrent{revsec}{\thesection}%
\zref@setcurrent{revpage}{\thepage}%
\zref@setcurrent{revcontent}{#2}%
\refstepcounter{revc}%
\label{#1}%
\zlabel{#1}%
\textcolor{blue}{#2}%
}

\newcommand{\revinu}[2]{%
\zref@setcurrent{revsec}{\thesection}%
\zref@setcurrent{revcontent}{#2}%
\refstepcounter{revc}%
\zlabel{#1}%
\label{#1}
#2 }

\newcommand{\revr}[2]{%
\zref@setcurrent{revsec}{\thesection}%
\zref@setcurrent{revcontent}{#2}%
\refstepcounter{revc}%
\zlabel{#1}%
\label{#1} \sot{#2}} \makeatother

\expandafter\def\expandafter\quote\expandafter{\quote\onehalfspacing\fontsize{12}{14}\selectfont}

\usepackage[framemethod=tikz]{mdframed}
\usepackage{lipsum}

\definecolor{mycolor}{rgb}{0.122, 0.435, 0.698}

\newmdenv[innerlinewidth=0.5pt, roundcorner=4pt,linecolor=mycolor,innerleftmargin=6pt,
innerrightmargin=6pt,innertopmargin=6pt,innerbottommargin=6pt]{mybox}

\begin{document}
\sloppy

\title{Game-Theoretic Mode Scheduling  for\\ Dynamic TDD in 5G  Systems }
\author{Chandan Kumar Sheemar, {\em Student Member, IEEE}, Leonardo Badia, {\em Senior Member, IEEE}, \\ and Stefano Tomasin {\em Senior Member, IEEE} \thanks{C. K. Sheemar is with EURECOM, Sophia-Antipolis, France, email: sheemar@eurecom.fr.
L. Badia and S. Tomasin are with the Department of Information Engineering, University of Padova, Italy, emails: badia@dei.unipd.it and tomasin@dei.unipd.it. L. Badia and S. Tomasin are also with the Italian National Consortium for Telecommunications (CNIT). This work has been in part supported by  MIUR (Italian Ministry of Education, University, and Research) under the initiative ``Departments of Excellence'' (Law 232/2016).
}}
\maketitle
\begin{abstract}
Dynamic time-division duplexing (TDD) enables independent uplink/downlink mode scheduling  at each cell, based on the local traffic. However, this creates cross-interference among cells. Thus, the joint power allocation and scheduling problem becomes mixed-integer non-convex and turns out to be NP-hard. We propose a low-complexity and decentralized solution, where power allocation and scheduling are decoupled. First,  power is allocated in a decentralized fashion, and then modes are scheduled by a non-cooperative game to achieve the mixed-strategy Nash equilibrium. We consider two possible approaches to compute the payoffs in the game, according to the cross-interference power model and the entailed communication overhead among cells. Simulation results are presented for an outdoor dense small-cell scenario, showing that our approaches outperform static TDD significantly. 
\end{abstract}
\IEEEpeerreviewmaketitle
 
\section{Introduction} \label{intro}

In cellular networks, uplink (UL) and downlink (DL) transmissions occur either at different times or over different frequency bands. In particular, with time division duplexing (TDD), the base station alternates time periods when it receives  signals from its served users in the UL, or transmits to users in the DL. In advanced long-term evolution (A-LTE) networks, a fixed configuration of the TDD periods is used, to be chosen among a given set. In the new radio (NR) standard for fifth-generation (5G) networks, this rigid scheme is broken, and \emph{dynamic TDD} (DTDD) \cite{3gpp-38213-f50}  enables cells to independently set the UL/DL configuration in each slot.

DTDD introduces UL-to-DL and DL-to-UL cross-interference among cells, thus, suitable cross-interference management techniques must be adopted, starting from a proper UL/DL mode scheduling.  Recent solutions for centralized scheduling include interference cancellation techniques \cite{8611365} and use of full-duplex devices \cite{razlighi2020optimal}.
Decentralized solutions are available in \cite{dowhuszko2013decentralized,elbamby2014dynamic,venkatasubramanian2014performance,shen2012dynamic}. In \cite{dowhuszko2013decentralized}, a frame of 10~ms is split into a DL and a UL period, and the switching points between the two periods is defined independently by each small cell, based on exchanged information. In \cite{elbamby2014dynamic}, the switching point decision problem is formulated as a non-cooperative game, to minimize the overall UL and DL delay at each cell. In \cite{venkatasubramanian2014performance}, the performance gain of centralized and decentralized DTDD is studied and compared with TDD. In \cite{shen2012dynamic}, the authors present a decentralized approach by grouping strongly coupled cells into clusters. In NR, to fully exploit the flexibility offered by DTDD, every cell should be able to choose scheduling modes at every slot independently. However,  state-of-the-art solutions considerably restrict the potential flexibility of DTDD by considering only two UL/DL periods per frame, or by clustering cells.  Moreover, for a better scheduling, also the transmit power at each cell must be adapted to the observed overall interference. However, in the existing literature, power is not optimized together with the scheduling pattern.  

In this letter, we consider the joint scheduling and power allocation problem, aiming at a maximizing the average weighted sum-rate (WSR) of the network: this is mixed-integer and non-convex (MINC), which turns out to be NP-hard \cite{burer2012non}. Thus, we decouple the  problem: first, power is allocated with a low complexity, greedy, and decentralized algorithm, then, the mode is scheduled by playing a game among cells. We consider a non-cooperative game, wherein  each base station plays an action (the UL or DL mode) with a given probability distribution, so as to maximize its utility function, given the choices made by its opponents. The neighboring cells' choices are crucial for the resulting utility, due to cross-interference. The game is solved  according to the mixed-strategy Nash equilibrium (MS-NE) concept, which always exists for finite games \cite{osborne}. We consider two payoffs, using different models for the cross-interference and requiring different overheads for the information exchange among cells. The proposed scheduling is performed at each slot independently, a solution that can be included in future releases of NR, as currently only a limited number of slot mode configurations can be selected.

Simulation results are presented for a dense, outdoor, and small-cells scenario, with line-of-sight (LoS) links. To our best knowledge, this is the first contribution to provide results for the most challenging DTDD scenario of \cite{shen2012dynamic}, without resorting to clustering.

In summary, the contributions of this letter are:
\begin{itemize}
    \item the definition of the joint scheduling and power allocation problem for DTDD  for average weighted sum-rate maximization;
    \item the decoupling of the power allocation and scheduling problem in DTDD; we obtain a much simpler system of (nonlinear) equations suited for a very efficient solution, without iterations or tuning of multiple parameters, as in \cite{dowhuszko2013decentralized};
    \item a novel scheduling procedure, implemented as a game among  cells;  the MS-NE solution is computed once over many slots, as long as the channel can be considered invariant, whereas DTDD modes are still selected for each slot. This leads to a significant saving of computational resources, while exploiting the full flexibility of DTDD.
    \item the definition of a simplified payoff model for the scheduling game that reduces the information exchange overhead among cells.
\end{itemize}
Lastly, note that we do not cluster cells for DTDD scheduling, which allows each cell to independently adapt the DTDD modes to its local traffic.

\textbf{Notation:} $\mathbb{P}[\cdot]$ and $\mathbb{E}_x[\cdot]$ denote probability and expectation (taken with respect to random variable $x$), respectively. Vectors and matrices are in boldface, while scalars are in italic. $(x)^+ = x$, when $x>0$ and $(x)^+ =0$ otherwise.
\section{System Model} \label{Section2}
 
We consider $N$ cells, where cell $n$, $n=1, \ldots, N$, serves users with indices in the set $\mathcal U_n$, and $|\mathcal U_n| = K , \forall n$. Thus, the total number of users is $NK$, which are identified by indices $k=1, \ldots, NK$. Time is split into slots. Each slot comprises multiple {\em resource blocks} of 12 subcarriers each. For the sake of a simpler explanation, we assume that the same number of simultaneous resource blocks is allocated to
each user for an entire slot.  Moreover, the $K$ users in a cell occupy all the available spectrum. We denote the resulting per-user portion of the time-frequency plane as {\em user block}.\footnote{This can be easily extended to more general cases of different
resources allocated to each user and different numbers of users per
cell.}
Let $ \phi_i \in \{1, \ldots, K\}$ denote the user block assigned to user $i \in \mathcal U_n , \forall n$. As only one user is allocated to each user block in each cell, $N$ users are served simultaneously in the network on the same user block.
 
In each slot, cell $n$ can be in either DL or UL mode, indicated  as $z_n =1$ or  $z_n =0$, respectively.  Moreover, at each slot, the UL and DL transmit powers are also set. Let $\bm{z} = [z_1, \ldots, z_N]$ denote the chosen modes for all cells at each slot. Let ${\tt p}_{k}(\bm{z})$ be the power allocated either to or from user  $k$, in DL or UL mode, respectively. Note that the dependence of power allocation arises due to cross-interference, which must be taken into account. We consider the following power constraints for  users and  base stations:
\begin{subequations}
\label{powconst}
\begin{equation}
 {\tt p}_{k}  \leq \alpha_0, \quad k \in \mathcal{U}_n, \mbox{ if $z_n=0$},
 \end{equation}
 \begin{equation}
  \label{powerDL}
  \sum_{k \in \mathcal{U}_n}    {\tt p}_{k} \leq \alpha_1,  \mbox{ if $z_n=1$},
\end{equation}
with mode-dependent constants $\alpha_0$ and  $\alpha_1$, respectively. \end{subequations}

We assume that the link between user $k$ and its serving base station $n$ is a narrowband channel with gain $H_{n,k}$ and we let $\bm{H}$ be the $N \times NK$ matrix with entries $\{H_{n,k}\}$. Moreover, since  users operating in UL mode interfere with the DL users of neighboring cells (UL-to-DL interference), we define the $NK \times NK$ symmetric matrix $ \bm{U}$, whose entry $U_{k_1,k_2}$,  $k_1, k_2 = 1, \ldots, NK$, is the  channel gain   between users $k_1$ and $k_2$. Similarly,  base stations operating in DL interfere with base stations operating in UL (DL-to-UL interference), thus we define the $N \times N $ symmetric matrix $\bm{B}$, whose entry $B_{n_1, n_2}$, $n_1, n_2 = 1, \ldots, N$, is the  channel gain    between base stations $n_1$ and $n_2$. We assume also that channel reciprocity holds for $\bm{H},\bm{U}$, and $\bm{B}$.

\subsection{Scheduling and Power Allocation Problem} 
 
To compute the network WSR, used as optimization metric, let the power of interference suffered by user $k$ served in cell $n$ operating in DL mode ($z_n = 1$) be
\begin{equation} 
    I_{n,k} (\bm{z}) =    \sum_{m =1, m\neq n }^N I'_{n,k,m}(\bm{z}) ,
\label{sumint}  
\end{equation}
\begin{equation} 
    I'_{n,k,m}  (\bm{z}) =       z_m H_{m,k} \hspace{-5mm} \sum\limits_{\substack{j \in \mathcal{U}_m: \phi_j=\phi_k }}  \hspace{-5mm}  {\tt p}_{j}   
     + c_1   (1 - z_m) \hspace{-5mm} \sum\limits_{\substack{j \in \mathcal{U}_m: \phi_j=\phi_k }}   \hspace{-5mm}  U_{j,k} {\tt p}_{j},
    \label{int1}
\end{equation}
with  $c_1 \in [0,1]$ being the UL-to-DL   cross-interference management factor.  
In (\ref{int1}), we account for the interference from other base stations operating in DL (matrix $\bm{H}$) and other users operating in UL (matrix $\bm{U}$). Similarly,  the power of interference suffered by user $k$ served in cell $n$ operating in UL mode ($z_n = 0$) is given by  \eqref{sumint}, now with 
\begin{equation}
  I'_{n,k,m} (\bm{z}) =  c_2     z_m B_{m,n}  \hspace{-5mm}\sum\limits_{\substack{j \in \mathcal{U}_m: \phi_j=\phi_k}}  \hspace{-5mm} {\tt p}_{j}   +   (1-z_m)
\hspace{-5mm} \sum\limits_{\substack{j \in \mathcal{U}_m:\phi_j=\phi_k }} \hspace{-5mm}  H_{n,j} {\tt p}_{j},
 \label{int2}
 \end{equation}
 where   $c_2 \in [0,1]$ is the DL-to-UL cross-interference management factor. In (\ref{int2}), we  take into account the interference from other base stations operating in DL (matrix $\bm{B}$) and  other users operating in UL (matrix $\bm{H}$). 
 Note that $c_1$ and $c_2$ varying in the interval $[0,1]$ assume value $0$ or $1$ in the case of perfect or no interference management, respectively.
 For a given set of modes $\bm{z}$, the WSR at base station $n$ is then
\begin{equation}
\begin{split}
W_n & (\bm{z}) = \sum_{k \in {\mathcal U}_n}  w_k(z_n)  \log_2 \left(  1 +   \frac{  H_{n,k}  {\tt p}_{k}}{\sigma^2 +  I_{n,k}(\bm{z}) }\right),
\end{split}
\label{MAXEQUATION}
\end{equation}
where $\sigma^2$ is the variance of the noise, $w_k(z_n) \in [0,1]$ is the weight of user $k \in \mathcal U_n$, when the serving base station $n$ is operating in mode $z_n$. The network WSR is
\begin{equation} 
W(\bm{z}) = \sum_{n = 1}^{N} W_n(\bm{z}). \label{global}
\end{equation}
 
A possible target for scheduling optimization is the maximization of the instantaneous WSR, i.e.,
\begin{equation}
\begin{aligned}
    &\max_{\{{\tt p}_{k}(\bm{z})\}, \bm{z}}  
     W(\bm{z}), \mbox{s.t. (\ref{powconst}) and $z_n \in \{0,1\}$}, \forall n.\\
    \end{aligned}\label{optimproblem}
\end{equation}
However, the joint scheduling and power allocation problem (\ref{optimproblem}) is  MINC  and turns out to be NP-hard \cite{burer2012non}. In particular, its non-convexity comes from the interference terms. Note that, to fully exploit DTDD, the problem  must be solved at each slot (millisecond scale), which is prohibitive.

\section{Scheduling and Power Allocation Decoupling}
 
To simplify and decentralize the problem, we assume that each base station randomly selects its own mode, thus $z_n$ becomes a Bernoulli random variable. Let $q_n$ be the probability that base station $n$ selects $z_n=1$, i.e.,
\begin{equation}
    q_{n} = \mathbb P[z_n = 1], \quad 1-q_{n} = \mathbb P[z_n = 0].   \label{defqn} 
\end{equation}
The value of $q_n$ is chosen to maximize the {\em average} WSR, i.e.,
\begin{equation} 
    \max_{q_n} \mathbb{E}_{\bm{z}}\left[{W_n(\bm{z})}\right]\,,
    \label{maxprobl}
\end{equation}
with the expectation taken over the actions of all players. Note that problem \eqref{maxprobl} should be solved once over multiple slots, as long as channel conditions and rate weights do not change, thus the network WSR for a given mode configuration does not change. Still, the mode is chosen for each slot according to the obtained distribution $q_n$. 

Problem \eqref{maxprobl} requires the computation of $W_n(\bm{z})$, which, from \eqref{MAXEQUATION} requires the knowledge of all channel gains. To reduce the communication overhead needed to share this information among all base stations, we approximate the instantaneous interference power level with its average, as better detailed in the following. 

\subsection{Average Interference Power}

We define the average interference power for cell $n$ as
\begin{equation}
    J_n(\bm{z}) =      \sum_{m =1, m\neq n }^N \bar{I}_{n,m}(\bm{z}),
    \label{ibaruno}
\end{equation}
where $\bar{I}_{n,m} (\bm{z})$ is the average interference power suffered by users in cell $n$ from cell $m$, i.e.,
\begin{equation}
\begin{split}
   \bar{I}_{n,m} (\bm{z}) = &  \mathbb{E}_{\bm{B}, \bm{H}, \bm{U}, \{\tt p_k\}}[I_{n,k,m}'(\bm{z})],
  \label{intmulK_1}
\end{split}
\end{equation}
and the expectation is taken with respect to both the channel gains and the powers.

The computation of the average in \eqref{intmulK_1} is rather complex, since the allocated power depends on scheduling and power assignments, which in turn depend on the channel gains. Thus, we consider an approximated value of the average interference power, where we keep the transmit power as a fixed term. In particular, we assume the DL transmit power as equally split (on average)  among the $K$ served users, i.e., ${\tt p}_{k} = \alpha_1/K$ for $k \in \mathcal U_{m}$ when $z_{m}=1$, $\forall m$. In UL, we assume that all users transmit at their maximum power, i.e., ${\tt p}_{k} = \alpha_0$, for $k \in \mathcal U_{m}$ when $z_{m}=0$, $\forall m$. Then, under this setting, the average interference power
suffered by users in cell $n$ from cell $m$ becomes
\begin{equation}
\begin{split}
   \bar{I}_{n,m} (\bm{z}) = &   \mathbb{E}_{\bm{B}, \bm{H}, \bm{U}}[I_{n,k,m}'(\bm{z})],  
  \label{intmulK}
\end{split}
 \end{equation}
while \eqref{ibaruno} still holds.
When cell $n$ is in DL ($z_n=0$),  since we have one user per user block from \eqref{int1} we have 
\begin{equation}
\begin{split}
   \bar{I}_{n,m} (\bm{z}) = &   \frac{\alpha_1}{K}  z_m\mathbb{E}_{ \bm{H} }[H_{m,k}]       + c_1   (1 - z_m)   \mathbb{E}_{\bm{U}}[U_{j,k}] \alpha_0,
  \label{int1m}
\end{split}
 \end{equation}
 while when cell $n$ is in UL ($z_n=1$), from \eqref{int2}  we have 
\begin{equation}
\begin{split}
   \bar{I}_{n,m} (\bm{z}) = &    \frac{\alpha_1}{K}c_2     z_m\mathbb{E}_{\bm{B}}[B_{m,n}]      +   (1-z_m)
  \mathbb{E}_{\bm{H}}[H_{n,j}] \alpha_0.
  \label{int2m}
\end{split}
 \end{equation}
Note that the expectations here can be computed either using the channel statistics, if available, or by averaging multiple channel estimates over time. When using this approximation $J_n(\bm{z})$ in the following will be denoted as  {\em reference interference power}.

\subsection{Power Allocation Sub-problem} 
 For power allocation, we assume that each cell allocates power to maximize only its utility function, regardless of  interference and cross-interference. This enables a greedy, decentralized, and low complexity power allocation.  
 
In UL mode, we assume that  each user transmits at its maximum power, to maximize its rate, i.e., 

\begin{equation}
    {\tt p}_k^* = \alpha_0 \quad k \in \mathcal{U}_n, \mbox{ if $z_n=0$}. \label{eq1}
\end{equation}
 
For the DL mode, we approximate $I_{n,k}(\bm{z}) \approx  J_n(\bm{z})$ and we distribute the transmit power according to a modified version of  the (weighted) water-filling algorithm  \cite{Tse-book}, which takes into account the noise, interference, and cross-interference and aims at maximizing $W_n^s(\bm{z})$ at each cell independently, i.e.,
\begin{equation}
    {\tt p}_k^* = \left(\frac{w_k(1)}{\lambda_k} - \frac{\sigma^2 + J_{n}(\bm{z})}{H_{n,k}} \right)^+, \mbox{ if $z_n=1$},\label{eq2} 
\end{equation}
where $\lambda_k >0$ is the Lagrange multiplier that satisfies the power constraint (\ref{powerDL}).

\subsection{Approximated WSRs}

Once powers have been allocated, let  $ W^{\tt s}_n(\bm{z})$  be the WSR (\ref{MAXEQUATION}), with powers given either by  (\ref{eq1}) or (\ref{eq2}), depending on the DTDD mode, to be used in \eqref{maxprobl}.

A further simplification is obtained by replacing $I_{n,k}(\bm{z})$ with $J_{n}(\bm{z})$ in (\ref{MAXEQUATION}), providing the WSR 
$W_n^{\tt a}(\bm{z})$, which is  then used instead of  
$W_n^{\tt s}(\bm{z})$    
 in \eqref{maxprobl}.
 
\section{Game-theoretic Scheduling}\label{static_sec} 

As the major contribution  of  this  letter, we tackle this issue by framing  the scheduling  problem  \eqref{maxprobl} as a  non-cooperative, decentralized, and instantaneous game,  played  by  the  base  stations.   The MS-NE solution of the game provides an efficient distributed scheduling, where no players desire to deviate from.
Differently from a centralized scheduling scheme, each base station $n$ acts as a player on its own, with action set $\mathcal{A}_n = \{0,1\}$, corresponding to  UL and DL modes.  Let the overall set of actions be $\mathcal{A}= \mathcal{A}_1\times \ldots\times \mathcal{A}_N$. For each slot, each base station $n$ chooses its mode over its action set $\mathcal{A}_n$, according to the distribution $q_n$. Define $\bm{q} = [q_1,\ldots, q_N]^T$ as the vector of probabilities of choosing the DL mode for the $N$ base stations: this is the vector of mixed strategies over the action set $\mathcal{A}$.  Since we aim at maximizing the average WSR, the payoff of player $n$, denoted as  $W_n(\bm{z})$, can be either the WSR $W^{\tt s}_n(\bm{z})$ or its approximated version  $W^{\tt a}_n(\bm{z})$. The objective of each player is the maximization of his/her average payoff, through the choice of a mixed strategy.

Now, the scheduling problem translates into the following static game (in normal form) with players, actions, and payoffs
\begin{equation}
\mathcal{G}  = \left(\{1, \ldots, N\},\mathcal{A}, \{{W_n(\bm{z})}\}  \right).
\end{equation}

\subsection{ Mixed Strategy Selection}  \label{mixed strategy}

We find an MS-NE, where each player aims at maximizing its average payoff (solution of \eqref{maxprobl}); if any player deviates, then its payoff degrades.
In DTDD, this means that the equilibrium distribution locally maximizes the payoff at each cell. To solve the optimization problem in the MS-NE sense, we impose also the {\it principle of indifference} \cite{osborne}. This principle imposes the further constraint that player $n$ chooses $q_n$ also to make his/her opponents indifferent over their actions. More precisely, the opponents would obtain the same average payoff, no matter which action they choose. Hence, the  cross interference with optimal distribution yields that   the same rate in UL and DL is achieved by  neighboring cells.  

This is achieved only in the asymptotic regime, with many opportunities to alternate between UL and DL, while  in practice channel and rate weights can be considered constant only over a few  slots, yielding only few opportunities.

 For user $n$ in UL and DL modes, the average payoff is denoted as $\mathbb{E}_{\bm{z}}\big[ W_n(\bm{z})| z_n=a \big]$, with $a = 0$ or 1, respectively, where the expectation is taken with respect to all the possible actions $z_m$, $m\neq n$ of the opponents of player $n$. By imposing the indifference principle, we obtain the following system of $N$ nonlinear equations in $N$ variables ($q_n$)  [Sec. 4.6.2 \cite{osborne}]
\begin{equation}
\mathbb{E}_{\bm{z}}\big[ W_n(\bm{z})| z_n=1 \big] = \mathbb{E}_{\bm{z}}\big[ W_n (\bm{z})| z_n=0 \big], \; n{=}1,{\ldots}, N.
\label{Peq} 
\end{equation}
 The average payoff of player $n$, can be written as a function of the probability that the various modes are selected, i.e.,   [see (113.1) \cite{osborne}]
\begin{equation} 
\begin{split}
	\mathbb{E}_{\bm{z}} &\left[ W_n  (\bm{z}) \right | z_n=a] =  \hspace{-5mm} 
	\sum\limits_{\bm{z'} \in \mathcal{A}: z'_n = a } W_n(\bm{z}') {\mathbb P}[\bm{z} = \bm{z}'] \\
	& =
	\sum\limits_{\bm{z'} \in \mathcal{A}: z'_n = a } W_n(\bm{z}')  \hspace{-1mm} \prod_{\substack{m = 1, m \neq n\\}}^{N} \hspace{-3mm} q_m^{z'_{m}} (1-q_{m})^{1-z'_{m}}, \label{rate_DLUL}
\end{split}
\end{equation}
where the latter equality is obtained considering that modes are selected independently by each player $n$ with probabilities $q_n$ (see \eqref{defqn}).

 Given the information shared from neighboring cells, \eqref{Peq} and \eqref{rate_DLUL} provide a system of polynomial equations in $q_m \in [0,1]$, which can solved by each cell  using numerical methods (see \cite{TELEN2018119} and references therein)  to pick its mixed strategy. Moreover, each cell also knows the strategies of others as every player solve the same system \eqref{Peq}, according to MS-NE.

Problem (\ref{optimproblem}) is then shifted into a system of equations, which can be solved very efficiently. To further save computational complexity, \eqref{Peq} can be solved  only at one base station and the optimal distribution communicated to the others, as all players solve the same system. The amount of information shared strictly depends on the type of payoffs, as discussed in Section~\ref{payoff_model}. Lastly, note that the MS-NE system  of equations has at least one  solution as the actions set of DTDD players is finite [Prop. 116.1 \cite{osborne}]. However, being nonlinear, it may exhibit multiple solutions. In that case, all the players must agree to choose the mixed strategy that maximizes the overall system performance, to avoid ambiguity.

\subsection{Impact Of The Payoff Model} \label{payoff_model}
We now consider the different alternatives of payoff $W_n(\bm{z})$, and their consequences on the communication overhead.

\paragraph*{Simplified-payoff (SIP) game with payoff $W_n^{\tt s}(\bm{z})$} When $W_n(\bm{z}) = W_n^{\tt s}(\bm{z})$, the interference power $I_{n,k}(\bm{z})$ has to be computed. In this case, although players independently solve \eqref{Peq}, they need to share information on the mutual interference, in particular, matrices $\bm{B}$, $\bm{H}$, $\bm{U}$, weights $\{w_k(z_n)\}$, and the allocated powers.

\paragraph*{Approximated-payoff (APP) game with payoff $W_n^{\tt a}(\bm{z})$}  
When $W_n(\bm{z}) = W_n^{\tt a}(\bm{z})$, the base stations  only need to share information about weights $\{w_k(z_n)\}$, allocated powers and $NK$ elements of matrix $\bm{H}$, corresponding to the direct signal channel gains.  All the cross-interference channel gains instead can be omitted. Still, we need the average interference power $J_n(\bm{z})$. From (\ref{ibaruno}), we observe that this can be obtained from the $(N-1)$ average interference terms $\bar{I}_{n,m}(z_m)$, $m=1, \ldots, N$, $m\neq n$. Therefore, a significant reduction of signaling overhead is obtained with this payoff. Moreover, as these average interference terms are slowly time-varying, they can be shared at a slower rate that $\bm{H}$, making the signaling overhead negligible.

\section{Numerical Results} \label{risultati}  

We now evaluate the performance of both SIP and APP games in a DTDD 5G network. We assume that channel varies every $10$~ms (one frame) and each user block comprises one resource block of 12 subcarriers, continuously allocated to the same user for one DTDD frame. Therefore, only one computation of the MS-NE takes place to schedule $10$ slots.
For comparison purposes, we consider:

a)  the \emph{instantaneous optimal approach} (OPT), which works as follows. At each slot, we first compute the  WSR $W(\bm{z})$  of (\ref{global}) with $W_n^s(\bm{z})$ for all possible scheduling modes $\bm{z}$, using the power allocation of \eqref{eq1}-\eqref{eq2}, and then select the mode and powers yielding the maximum WSR. Note that we must evaluate $2^{N}$ scheduling modes (all possible $\bm{z}$ vectors of size $N$ with 0-1 entries). Moreover, the full channel state information must be transferred to a central node, where the WSRs are computed, and then the scheduling decision should be broadcast to all cells. These operations must be repeated at each slot;

 b) the static TDD (STDD) scheme with frame equally partitioned in both directions, i.e., $5$ slots in UL and $5$ in DL; 

 c) the decentralized scheme \cite{dowhuszko2013decentralized}, which always assigns the first $2$ and the last $2$ slots to DL and UL, respectively, whereas the rest of the slots are chosen by making the optimal allocation of a DL to UL switching point.

 In all cases, power is allocated in the greedy fashion, as in (\ref{eq1}) and (\ref{eq2}). The expectations of \eqref{int1m} and \eqref{int2m} have been computed by averaging multiple channel estimates.

 We consider an outdoor dense scenario, with seven  hexagonal small cells, having side of size $333$~m. LoS channels with path-loss exponent $2$ are assumed. This results to be the most challenging scenario for DTDD \cite{shen2012dynamic}.
We wrap the scenario into a sphere, which leads to each cell having $6$ interfering cells. We also assume $c_1 = c_2 = c$ and the maximum transmit powers at the base stations and users, $\alpha_1=\alpha_0=23$ dBm, as in \cite{elbamby2014dynamic}. Each base station and user is equipped with  one omni-directional antenna. The UL and DL rate weights are chosen from a uniform distribution. The noise power is -10 dB to the signal power.
 
 We remark that, although under the APP model the players use an approximation of the payoff,  we report here the resulting \emph{true} payoffs obtained with their actions, according to (\ref{MAXEQUATION}), by considering the actual cross-interference at each slot.
 
\begin{figure} 
	\centering
	\includegraphics[width=1\hsize]{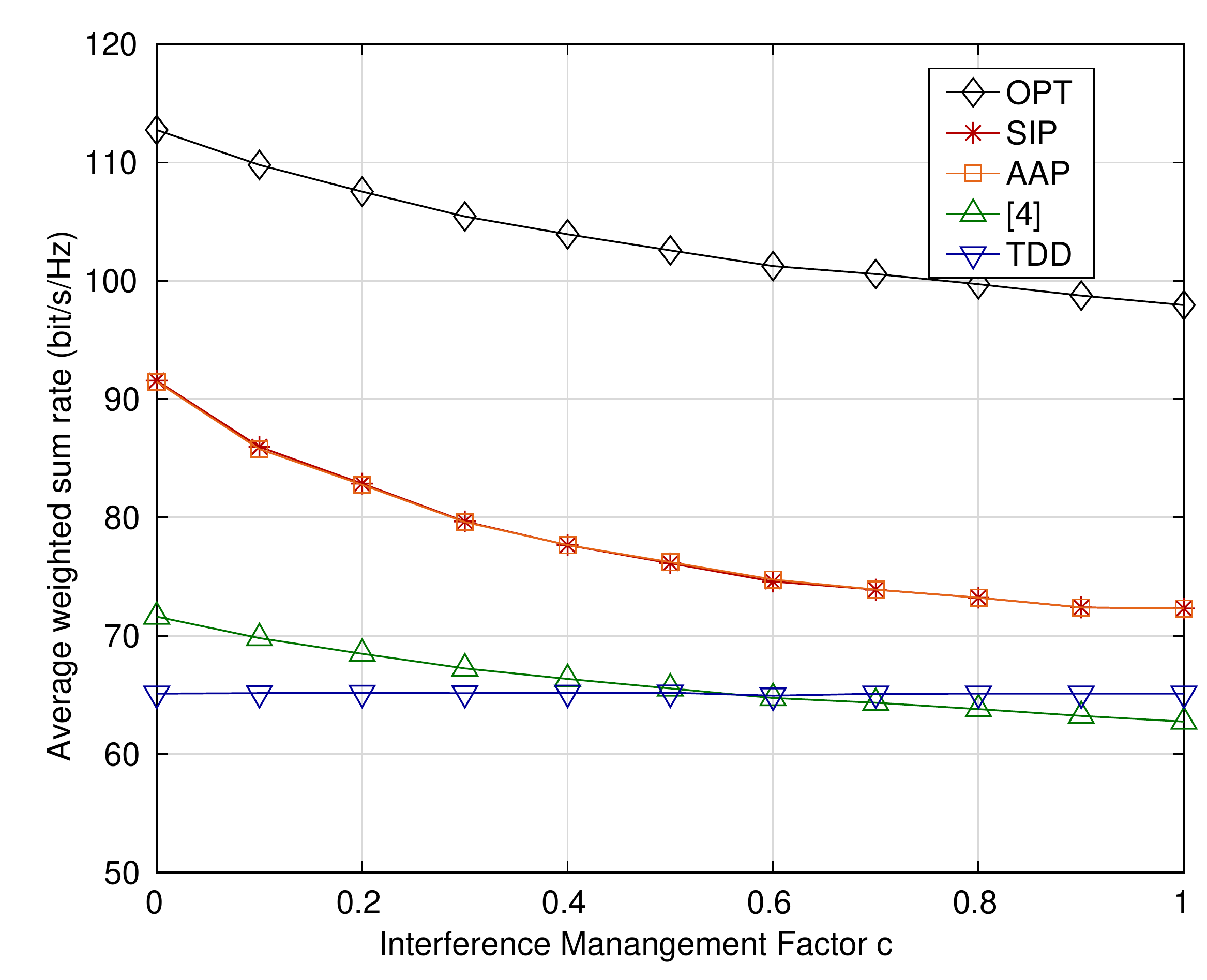}
 	\caption{Average network payoff of a DTDD system for OPT, SIP, APP, and STDD, as a function of $c$, with fixed 15 users per cell.}
 	\label{func_of_int}
\end{figure}

 Fig.~\ref{func_of_int} shows the average WSR achieved by the various approaches, as a function of $c$. We note that both SIP and AAP significantly outperform TDD and \cite{dowhuszko2013decentralized}, for all values of $c$. This is due to the higher flexibility of both SIP and AAP compared to TDD and  \cite{dowhuszko2013decentralized}. Moreover, APP performs strictly close to SIP, despite its use of the approximated payoff, which in turn makes APP quite appealing for its reduced communication overhead.  Still, SIP and APP exhibit a performance gap with respect to OPT, as in OPT we optimize the cell mode, thus maximizing the {\em instantaneous} sum-rate, while APP and SIP randomly select the cell modes, with densities that maximize the resulting {\em average} sum-rate. Moreover, with APP players do not have a perfect estimate of the interference when locally solving the MS-NE problem.

\begin{figure} %
	\centering
\includegraphics[width=1\hsize]{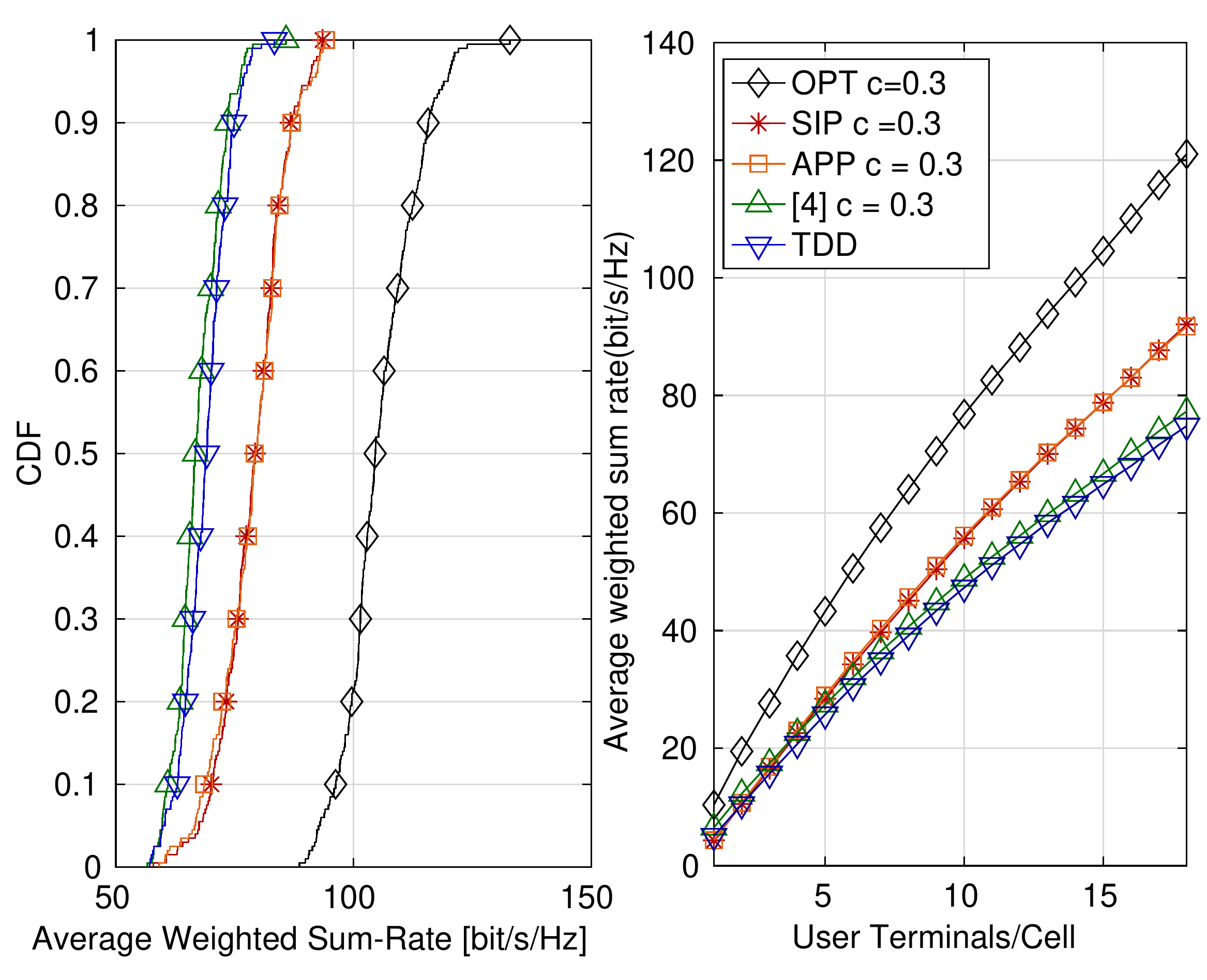}	
 	\caption{ a) CDF for the WSR with $c=0.3$ and $15$ users/cell. b) Average network payoff of a DTDD system for OPT, SIP, APP, and STDD, as a function of the number of users per cell.}
 	\label{func_of_users}
\end{figure}
 
  Fig.~\ref{func_of_users}.a shows the cumulative distribution function (CDF) of the average WSR for the various methods and typical cross-interference management factor  $c=0.3$. Both SIP and APP shows a similar CDF, and provides higher average rates than static TDD and the method of   \cite{dowhuszko2013decentralized}; this indicate the proposed solution operates fairly among cells. Fig.~\ref{func_of_users}.b shows the average WSR as a function of the number of users per cell, again for $c=0.3$.  According to our assumptions in the System Model, the user block shrinks in the frequency domain as the number of users increases. The figure  shows that our approaches are particularly effective in dense networks, which are of interest for current and future generations of cellular systems. Moreover, both games result to be extremely scalable as the network size grows. With decentralized cross-interference management, we provide a fully decentralized and low complexity DTDD scheduling for dense networks.

 \section{Conclusions} \label{conclusioni}
A new cross-interference aware, decentralized, and low complexity power allocation is proposed. Scheduling of UL/DL transmissions in a DTDD scenario is cast into the MS-NE of a simultaneous game among the base stations. Our performance evaluation shows a significant advantage with respect to TDD and the state-of-the-art in terms of achievable WSR even in absence of further interference rejection techniques for the most challenging DTDD scenario. In particular, the APP solution also significantly reduces the communication overhead, while performing close to SIP.

\bibliographystyle{IEEEtran}
\bibliography{biblio_letter}

 \end{document}